\documentclass[12pt]{iopart}
\usepackage{iopams}  
\usepackage{graphicx}

\begin{document}

\title[Kinematic measurements using an infrared sensor]{Kinematic measurements using an infrared sensor}

\author{F Marinho$^1$ and L Paulucci$^2$}
\address{$^1$ Universidade Federal do Rio de Janeiro, Av. Aluizio Gomes, 50, 27930-560, Maca\'e, RJ, Brazil}
\ead{marinho@macae.ufrj.br}
\address{$^2$ Universidade Federal do ABC, Rua Santa Ad\'elia, 166, 09210-170, Santo Andr\'e, SP, Brazil}

\date{\today}

\begin{abstract}
The use of an infrared sensor as a new alternative to measure position as a function of time in kinematic experiments was investigated using a microcontroller as data acquisition and control device. These are versatile sensors that offer advantages over the typical ultrasound devices. The setup described in this paper enables students to develop their own experiments promoting opportunities for learning physical concepts such as the different types of forces that can act on a body (gravitational, elastic, drag, etc.) and the resulting types of movements with good sensitivity within the $\rm 4-30~cm$ range. As proof of concept we also present the application of a prototype designed to record the kinematics of mass-spring systems. 
\end{abstract}

% Uncomment for PACS numbers
%
% Uncomment for keywords
\vspace{2pc}
\noindent{\it Keywords}: kinematics, infrared sensor, harmonic and damped oscillations, coupled springs.
%
% Uncomment for Submitted to journal title message
%\submitto{\JPA}
%
% Uncomment if a separate title page is required
%\maketitle
% 
% For two-column output uncomment the next line and choose [10pt] rather than [12pt] in the \documentclass declaration
%\ioptwocol

\section{Introduction}

The study of kinematics and movement dynamics is usually the first 
subject that students are formally introduced to in a physics classroom. 
The closeness to everyday situations indicates that these concepts are 
more easily understood than concepts in other branches of physics. 
Nevertheless it is often difficult to empirically measure the movement of bodies in a classroom environment or laboratory. Instead the measurement of the time interval between two previously determined positions or of the period for oscillating systems is usually the standard procedure in opposition to the direct measurement of the position, speed or acceleration of the body as a function of time.

Many approaches have been proposed in the literature for overcoming this 
and provide students with more detailed movement analysis. A traditional method for position determination as a function of time is to use a ticker-tape \cite{ticometro}. This method is useful for understanding the properties of uniform and uniformly accelerated movements and has been used worldwide. However, it is limited to single direction movements and measurements need to be read out with a ruler. More recently, the use of image recording devices \cite{image, coupled1}, gaming apparatuses \cite{games}, and other electronic pieces \cite{electronic} 
has gained attention. Also, with the popularisation of smartphones,
the use of their accelerometers \cite{acc} or light sensors \cite{light} 
was employed in mechanical systems analysis. However, in most of these 
applications, the need to develop specific software or to also use more 
elaborated (and sometimes expensive and large) apparatuses, such as 
air tracks or rotating platforms, was also necessary and it can pose 
some difficulties for its widespread use.

Using easily acquired and relatively low cost components, we present a 
setup that can make precise position measurements of a body in motion so 
that students can analyse the movement from the perspective of the 
position of a body at a given moment instead of the time interval it 
takes from moving between two fixed points. It employs infrared sensors 
from the SharpGP family replacing the typical sonars used in mechanical 
and oscillation experiments while for the acquisition, it uses an 
Arduino UNO board instead of a DAQ controller, thus offering simplicity 
from the technical point of view and also a better cost/benefit ratio. 
Commercial prices for a complete setup composed of sonar sensor, data 
acquisition interface, and software can cost over 500 USD while the 
setup proposed in this article should cost about 60 USD. 

The need to infer a minimum sample rate, the range 
of movement to be recorded, and the precision needed for the apparatus 
according to the investigation to be made can engage students in the 
experiment from the start, to discuss and come up with different 
strategies. Therefore, the setup phase of the prototype allows 
students to develop basic abilities for scientific instrumentation 
usually not present in traditional experimental physics classes. 

In this paper we discuss the applications of this method to record 
the oscillations of mass-spring systems, although further applications 
can be performed with few modifications. We show that the setup works 
very well for recording smooth motion. It, nevertheless, has not been 
tested for recording motion with sudden jumps in the acceleration.
Due to the averaging procedure used in section \ref{app}, a sudden
transition would be smeared into a sigmoid like transition when analysing the acceleration vs. time behaviour and a change in the regime would also be identifiable for the time-dependence of speed and position. In this way, it would be possible to recognise such condition in the data and, as usual in this kind of situation, it would be best to split the data analysis into two different acceleration regions.

\section{Experimental setup}\label{setup}

The proposed apparatus is mainly composed of a microcontroller and an 
infrared distance sensor as already mentioned. The distance sensor used 
was the Sharp GP2D120 analogue model \cite{sharp}. It is a sensor unit 
comprised of an infrared emitting diode, a position sensing detector, 
and signal processing electronics employing a triangulation method to
determine its distance to a target. Infrared radiation is emitted in a 
narrow beam which is reflected by the target back to the unit and readout by the position sensing detector. The position of the reflected beam on the detector surface is determined and the signal acquired is processed by the electronics which outputs a voltage that is dependent on the angle of the reflected beam, therefore, on the distance between the target and sensor. %This triangulation method is employed for better determination of distance rendering the influence from the ambient infrared contamination to be very much reduced.

Power and acquisition cables are 
connected to a microcontroller board programmed to record the sensor 
voltage output variation with time and transmit it to the computer via 
a USB port where data is readout and processed. 

The data acquisition and control platform used was an Arduino UNO
\cite{Arduino}, an open source device which presents special features 
for easily prototyping simple electronic setups. This choice is justified 
for its versatility, low cost, and friendly user software interface. It 
consists of a microcontroller ATMEL embedded in a board with a memory card 
and in/out peripherals. 

\subsection{Position calibration}\label{calib}

\begin{figure}[h!]
\centering
\includegraphics[width=0.5\textwidth]{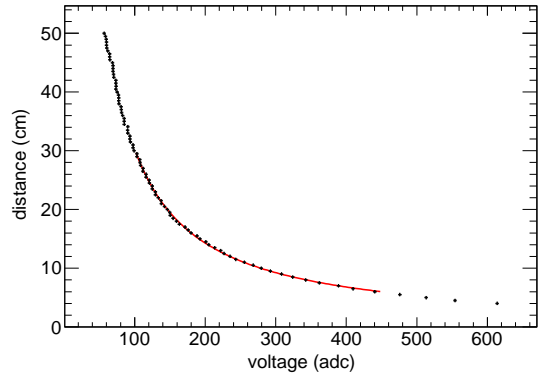}
\caption{Calibration curve of the infrared sensor. The distance from the sensor is shown as a function of the recorded output voltage. The red curve being the fit in the region of interest. Note the small error bars shown in the plot for both axes: the uncertainty for the position is 0.05 cm and for the voltage all error bars are smaller than 0.3\% of the correspondent measured value.} \label{calibfig}
\end{figure}

The calibration curve is obtained by setting up a squared white target in front of the sensor and recording the corresponding output voltage while varying the distance from the sensor, measured with a ruler.
The microcontroller was configured to register readings during five seconds for each position with an acquisition rate of 57600 bps. A capacitor, $C = 22 \mathrm{\mu F}$, was introduced between the 5V and ground terminals in the circuit for stabilising the sensor's energy supply.
The resulting curve is presented in figure \ref{calibfig}. 

In order to have a correspondence between the output voltage and distance between sensor and target, a hyperbolic equation, $d(V) = a + b~ V^{-c}$, was fitted to the data in the region of interest using ROOT \cite{root}. The calibration points and fitting curve can be seen in figure \ref{calibfig}. The parameters obtained were a=0.174(93), b=$456.7(13) \times 10^1$, and c=1.0909(67) for distance to be given in cm and the voltage in ADC units. The sensor's output gives better accuracy in the region between 4 and 30 cm. For distances above this value the sensor can still be used but the precision diminishes quickly the farther away the object is. For distances smaller than 3 cm the output read by the sensor increases with increasing distance so there is an ambiguity in determining the position, therefore such region must be avoided. The dispersion of the measurements was observed to vary from 2-9\% with respect to the mean value for distances of 4-35 cm, respectively.

Also, the reflectance of an object being dependent on material properties such as colour and surface texture can influence the output voltage recorded by the sensor, affecting the calibration. Nevertheless, as can be seen in Ref. \cite{sharp} for a comparison between white and grey papers, its influence is very small and mostly restricted to longer distances ($d > 20$ cm). Regarding the angle of incidence of radiation on the reflecting surface, the device is set to work better with reflection on a perpendicular surface. The more the surface angle deviates from it, the less reliable the measurement will be. Nevertheless it is not difficult to envision setups that can guarantee small deviations (if any) that will not appreciably affect the measurement. Such features should be taken into account when designing an application of the setup described here.

The sensor emits in the infrared radiation region with a wavelength of $\lambda =  850 (70)$ nm. The calibration presented was performed in a darken room with the whole apparatus protected from possible sources of bright lights, both natural and artificial. We have also taken some measurements without the protection to evaluate the influence of external noise. For a fixed position, we have evaluated the output read from the sensor in 3 situations: 

\begin{enumerate}
\item  setup blocked from external sources of light; 
\item setup mounted in a partially dark room with lights off and window's shutters closed, although some clarity still entered the room from outside; 
\item lights in the room on but avoiding any transitioning shades in front of the sensor and window's shutters closed. 
\end{enumerate}

The results measured for the same position were given by $V_1$ = 284.38 (26) adc, $V_2$ = 292.47 (40) adc, and $V_3$ = 292.34 (40) adc. The last two results are compatible with each other (with the artificial light negligibly contributing to a noise increase) but not with the first. Note that the standard uncertainty increased without the blockage of external light but the output read by the sensor increased by less than 3\%. It would imply in a mismatched position of only $\sim 3$\% off given the calibration curve obtained. Although the results are not compatible, the systematic error associated is low so that the prototype can be used even if it is not possible to cover the experimental setup. However, it is strongly recommended that measures to block direct sunlight are taken as this can be a time varying source of contamination for the measured signal.

\begin{figure}[h!]
\centering
\includegraphics[width=0.5\textwidth]{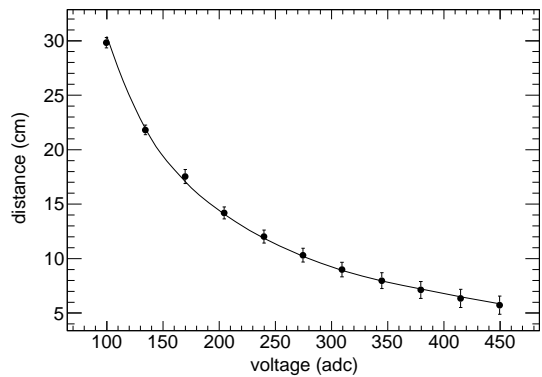}
\caption{Kinematics calibration curve of the infrared sensor. Data points are taken by recording the output voltage measured by Arduino of a moving object sliding on an air track. The full curve is the fit shown in figure \ref{calibfig} for comparison. Note that the two curves are compatible with each other and the uncertainties for this calibration are larger than those obtained for the static one.} \label{calibdyn}
\end{figure}

A kinematics calibration was also performed by attaching a white cardbox to an object sliding at an air track with a small inclination for obtaining the distance from the sensor as a function of the voltage recorded by it. The target was released at a certain distance from the sensor measured with a ruler to move towards it. The position of the target as a function of time was calculated with the kinematics equation for the uniformly accelerated motion. As can be seen in fig \ref{calibdyn}, the kinematics calibration is compatible with the static calibration. Nevertheless, the error bars in the case of a kinematics calibration are greater
due to the large uncertainty introduced by the synchronisation of the
starting time of data acquisition by the Arduino and the release of the sliding object. The uncertainty related to the angle of inclination is also a major limiting element for a precise position determination. Because of these factors it is overall difficult to calculate with good precision the position at a certain time and relate it to the voltage output measured with the Arduino without a complicated procedure to time align both quantities. Being a simpler and more precise procedure compatible with a dynamic calibration, the static one can be used without major concerns.

\begin{figure}[h!]
\centering
\includegraphics[width=0.5\textwidth]{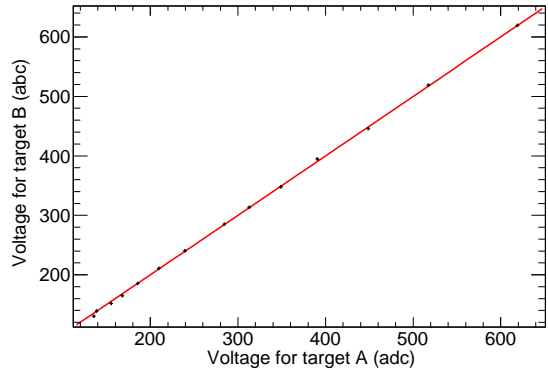}
\caption{Comparison between the voltage output recorded for two different sizes of reflecting objects. Data was taken using a 15 cm x 15 cm white cardboard (target A) and a 5 cm x 5 cm white cardboard (target B). Each point corresponds to the measurements of output voltage with both targets placed at the same distance from the sensor. The red curve shows the expected behaviour for same voltage, i.e., $V_A=V_B$ clearly showing the compatibility of both measurements.} \label{calibcomp}
\end{figure}

Although the setup described using arduino could possibly be adapted for use with a position sensor using different techniques, such as sonars, the use of an infrared sensor presents some advantages. Those include low cost, low power consumption, reduced dimensions, and very simple interfacing with readout electronics. Moreover since the wavelength emitted is of order of hundreds of nanometres (for the specific sensor used, it is given by 850 (70) nm), the size of the reflecting object is irrelevant, as long as it is aligned with the sensor's centre. However, limitations to align a moving target in front of the sensor put more stringent requirements on the target size. In order to guarantee an easy operation of the system, it is enough for the target to have dimensions larger than the distance between the sensor's emitter and receiver (in this case, $~2$ cm). We have used different sizes of target (15 cm x 15 cm and 5 cm x 5 cm) for the calibrations and the results are compatible with each other (see fig \ref{calibcomp}). This was expected since this is a triangulation sensor based on a position sensing detector with a narrow infrared beam. Therefore as long as the reflecting surface ensures enough beam irradiance on the sensor surface the unit is triggered.

\subsection{Data sampling rate}

A characterisation of the microcontroller used in the acquisition was 
performed for timing information and an overall evaluation of the data 
acquisition and transfer. The time Arduino takes to register two consecutive 
measurements can be easily obtained by counting the number of 
readings recorded by it during a fixed time interval (for example,
data taken in section \ref{calib} for position calibration could be used).

We have also tested the dependence of acquisition with different computers having different processing capabilities. The difference in the acquisition times were of the order of a few microseconds at 9600 bps data transfer rate, less than 1\% of the resulting time interval. These results indicate that using the expected acquisition time for a given data transfer rate, without performing this data sampling calibration, should not introduce significant uncertainties in the final result.

\begin{figure}[h!]
\centering
\includegraphics[width=0.5\textwidth]{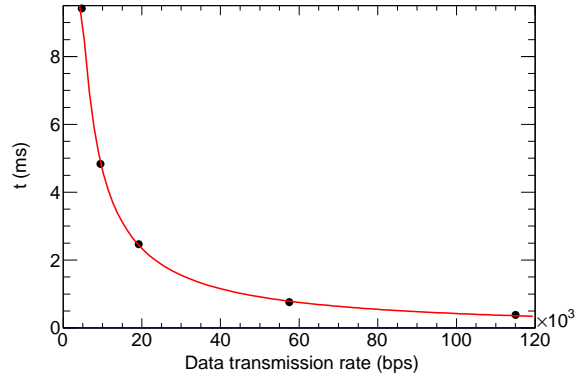}
\caption{Data sampling time interval for Arduino as a function of the data transmission rate. The best fit to the data is shown in red.} \label{ratefig}
\end{figure}

In figure \ref{ratefig} the obtained sampling time versus the data transmission rate is shown. As expected, the curve has a hyperbolic shape showing that the sampling time is determined solely by the data transfer rate and not limited by any other component of the acquisition system such as memory accessing, sensor dead time, etc. 

A maximum speed of about 10 m/s for the target is recommended to obtain good position determination at the least precise region close to the 30 cm distance to the sensor. This value is obtained considering the measurements dispersion, the maximum acquisition rate and the averaging procedure for neighbouring points described in this article. This speed value is one order of magnitude higher than the typical speeds recorded in the experiments of this article.

\section{Application to mass-spring systems}\label{app}

\subsection{Simple harmonic oscillator}

\begin{figure}[h!]
\centering
\includegraphics[width=0.4\textwidth]{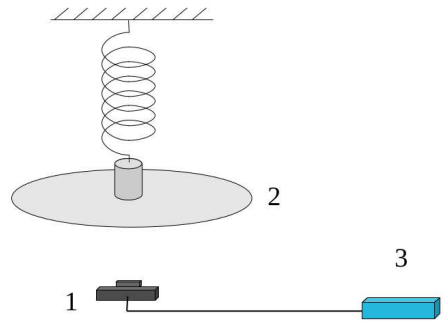}
\caption{Schematic setup for kinematic measurements of a harmonic oscillator. The IR sensor (1) emits infrared radiation which is reflected by the lower mass attached to the spring (2), here shown with a damping disk, and returns to the sensor allowing it to be read and transmitted to the Arduino board (3).} \label{springfig}
\end{figure}

The prototype was used to measure oscillations in a mass-spring system. Figure \ref{springfig} shows the schematic view of the setup used with its main components. 
The spring is fixed at one end and let to hang in the vertical direction while the mass is attached to its free end. The IR sensor is put directly below the mass and connected to the Arduino board. The  system is put to oscillate either by compressing or distending the spring in the vertical direction and the measured output can be transmitted to the computer. 

For an oscillating system of mass m and spring constant k with damping, 
assuming small oscillation amplitude and that the force acting against the 
movement is linearly dependent on the speed of the moving body (-bv), the 
movement equation describing the system can be written as \cite{oscilacao}
 			
\begin{equation}
m \frac{d^2y}{dt^2}+b\frac{dy}{dt}+ky=0,
\end{equation}
\noindent
with solution for weak damping given by
							
\begin{equation}
y(t) = A e^{-\gamma t}\cos{(\omega^\prime t + \phi)}, \label{singleeq}
\end{equation}
\noindent
where $\gamma= b/2m$ is the damping coefficient, $\omega^\prime = \sqrt{k/m -\gamma^2}$ is the system's angular frequency, and $\phi$ is the phase.  If b=0, the solution of the simple harmonic oscillator is recovered: $\omega = \sqrt{k/m}$.

The mass weight value used in the experiments was 95.00 (5) g with a 5 cm x 5 cm sheet of white paperboard attached to its bottom to reflect the emitted infrared signal. Two different springs were used while keeping the same weight. For the damped oscillator a large cardbox in the form of a disc of 26.00 (10) cm diameter was also attached to the system in order to increase the damping constant. The data was recorded with an acquisition rate of 9600 bps.

\begin{figure}[h!]
\centering
\includegraphics[width=0.5\textwidth]{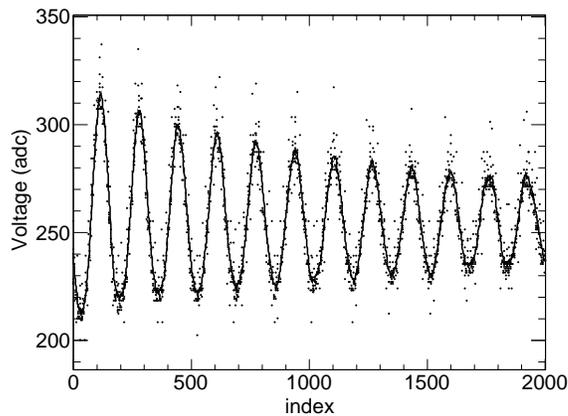}
\caption{Output measured during the oscillations of a damped harmonic oscillator. Raw values recorded by Arduino are shown as dots whereas a smoothing with 20 neighbouring points is shown as a solid line.} \label{averagefig}
\end{figure}

Figure \ref{averagefig} illustrates one data set recorded where the voltage output values are shown as a function of the number of measurement, indicated by the name index in the graph. Arduino sends to the computer a sequence of voltage values taken at equal time intervals. It means that the tenth measurement taken ($i=10$), for example, happened at a time $t=i~\Delta t$ after the run began, where $\Delta t$ is given by figure \ref{ratefig}. 
The dot markers in Figure \ref{averagefig} show the raw values for every single measurement. The solid line represents the average obtained with every set of neighbouring 20 points allowing drastic reduction of random noise contamination from the electronics. 

\begin{figure}[h!]
\centering
\includegraphics[width=0.49\textwidth]{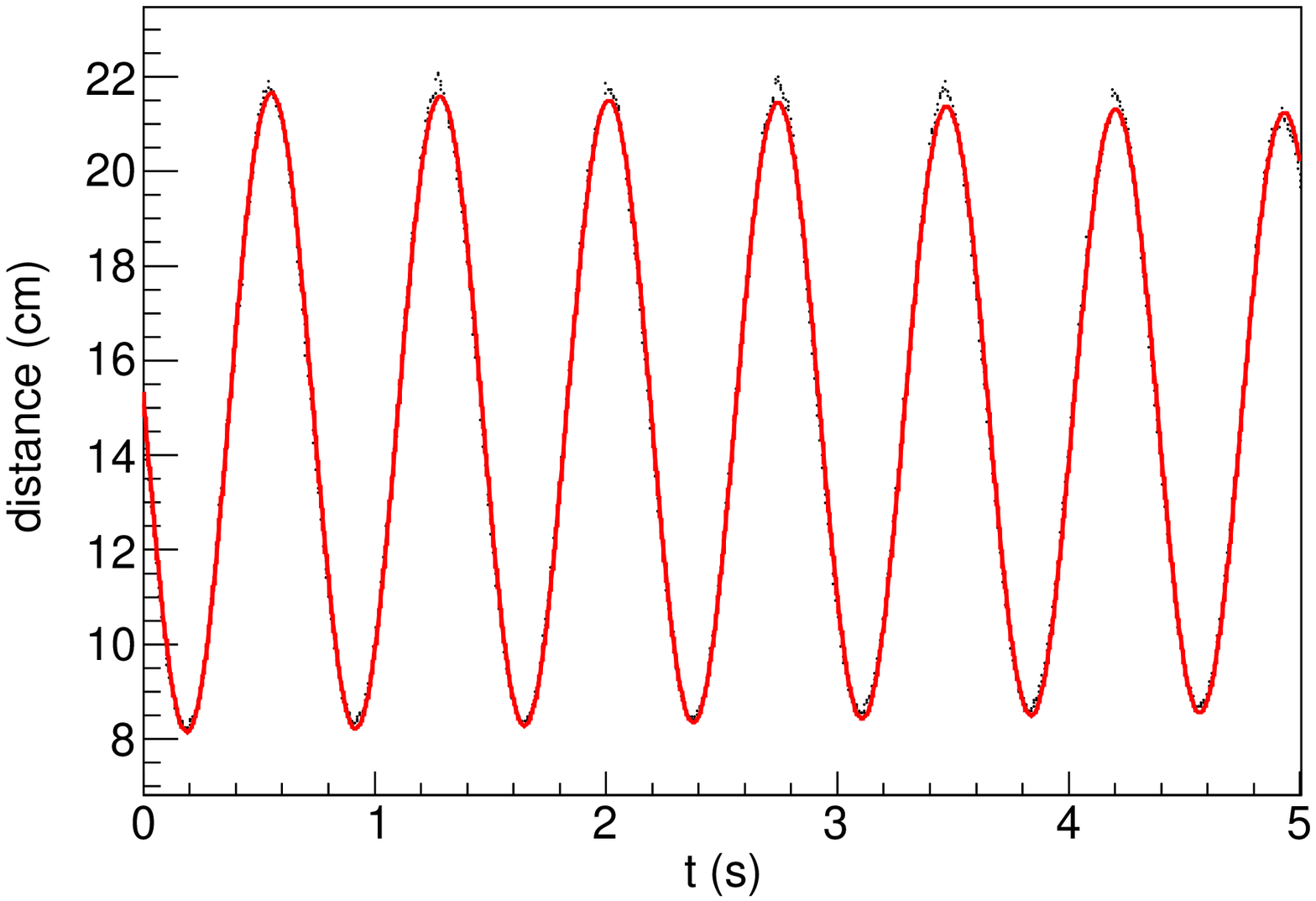}
\includegraphics[width=0.49\textwidth]{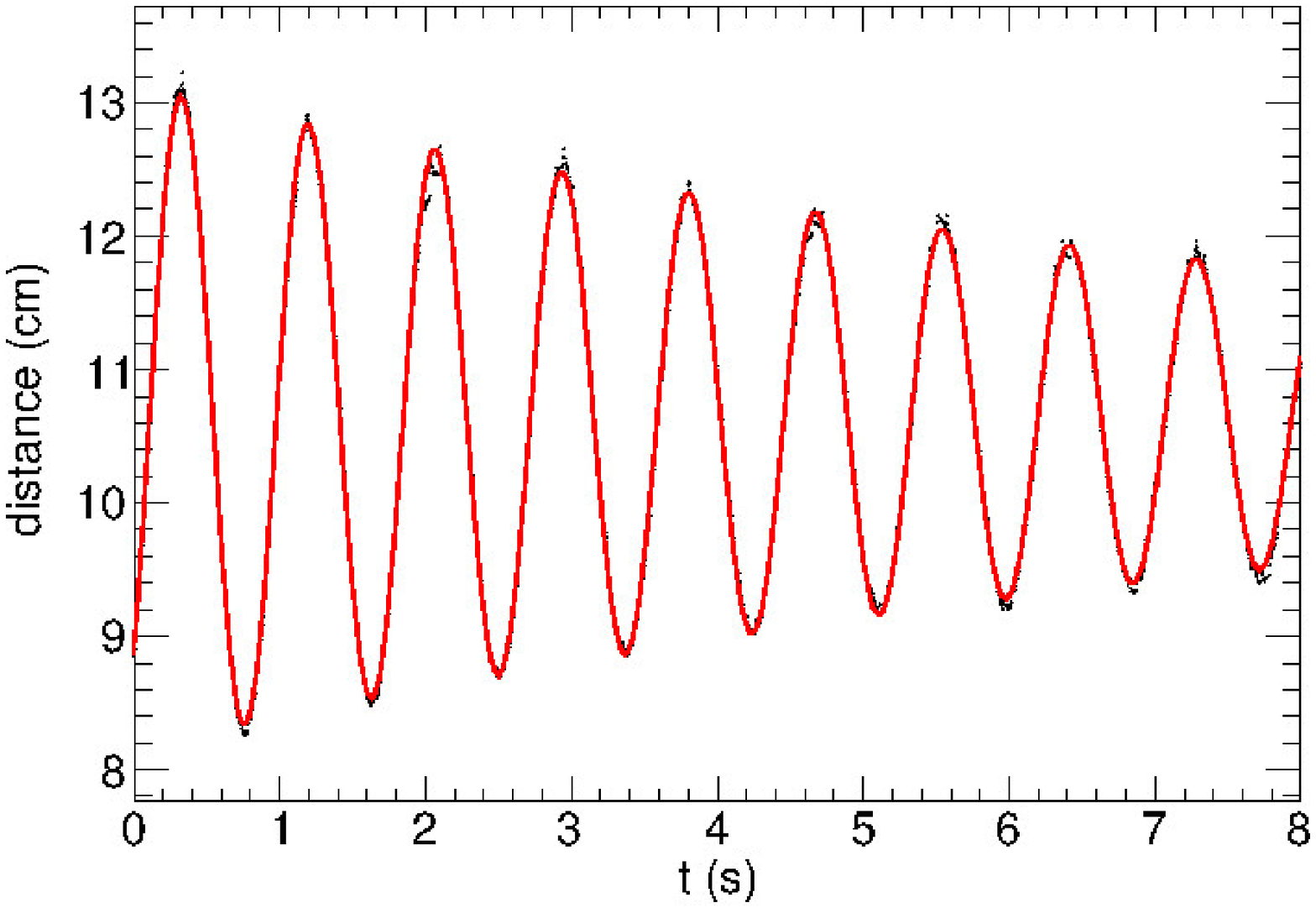}
\caption{Distance of the mass-spring system from the sensor as a function of time for two different damping coefficients but same mass weight and spring. On the left, a mass-spring system with negligible damping is shown while on the right the system is visibly damped. The same smoothing procedure shown in figure \ref{averagefig} was used here and is represented by the points in the graph while the curve is a fit of equation (\ref{singleeq}) to the data.} \label{singlefig}
\end{figure}

Data such as shown in figure \ref{averagefig} can be converted to the mass distance from the sensor as a function of time by using the position calibration obtained in section \ref{calib} and multiplying the index number by the adequate sampling interval shown in figure \ref{ratefig}. Examples of the resulting plots obtained from this procedure can be seen in figure \ref{singlefig}.

The converted data allows for the determination of the oscillation period, 
spring constant,
and damping constant by directly fitting equation 
(\ref{singleeq}) to the data. \footnote{A simpler way is to determine the period directly from the graph and, since the cosine function is limited, looking at equation (\ref{singleeq}) as having a time dependent amplitude, $A(t) = A_0 \exp(-\gamma t)$, and explore that to obtain the value of $\gamma$. It can be achieved with a linear fit to the logarithm of the time varying amplitude as a function of time taken at the consecutive maxima and minima.} The values of the two spring constants obtained were: $k_1=7.0422(48)$ N/m and $k_2=8.4342(73)$ N/m. It must be pointed out that it is also possible to fit the simple harmonic oscillator curve to the raw data. Therefore, it is not always necessary to resort to averaging, although this method makes the phenomenon much easier to be visualised.

\subsection{Coupled oscillators}

One interesting feature of oscillating systems that can also be shown with this apparatus is the coupling of springs. The theory of coupled oscillators is used to understand several physical phenomena such as vibration modes in solids, coupled oscillation systems in electronic circuits, etc. This is a type of oscillating system that is usually presented to students only through theoretical classes although some experimental visualisation of the phenomenon have been proposed \cite{coupled1, coupled2} showing interesting results. Nevertheless, once more, they mostly require the use of air tracks, air tables, development of image recognition software or electronic circuits to be performed. We propose a simple way of directly measuring the position in order to ease the students understanding of the phenomenon without requiring advanced knowledge in other branches of physics or programming skills.

\begin{figure}[h!]
\centering
\includegraphics[width=0.3\textwidth]{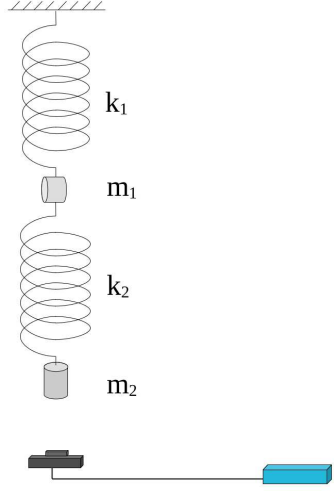}
\caption{Schematic view of the coupled springs system.} \label{coupledfig}
\end{figure}

For instance, figure \ref{coupledfig} shows a system composed of two springs put in the vertical direction. One spring had its upper end fixed while in the lower end a weight and another spring were attached. At the lower end of the second spring, another weight was put to oscillate and its distance from the sensor as a function of time was recorded. 

When attaching two ideal springs with two masses as seen in figure \ref{coupledfig}, for small oscillations, the following movement equations can be derived for the system \cite{oscilacao}:

\begin{eqnarray}
m_1 \frac{d^2 \Delta y_1}{dt^2} + \Delta y_1(k_1+k_2) &-& \Delta y_2 k_2  =  0, \\
m_2 \frac{d^2 \Delta y_2}{dt^2} +\,\,\,\,\,\, \Delta y_1 k_2 &+& \Delta y_2 k_2  =  0,
\end{eqnarray}
\noindent
with $\Delta y_j$ being the particles' displacements measured from their respective equilibrium positions. The above equations can be solved by assuming solutions of an oscillatory type: $\Delta y_j = C_j \exp{(i \omega t)}$, with $C_j$ constant. The two fundamental frequencies are found to be:
 					
\begin{equation}
\omega^2_{\pm} = \frac{(k_1+k_2)m_2+m_1 k_2}{2m_1 m_2} \Big[ 1 \pm \sqrt{1-\frac{4 m_1 m_2 k_1 k_2}{[(k_1+k_2) m_2+m_1 k_2]^2}}\Big]. \label{omegacoup}
\end{equation}
	 
\begin{figure}[h!]
\centering
\includegraphics[width=0.49\textwidth]{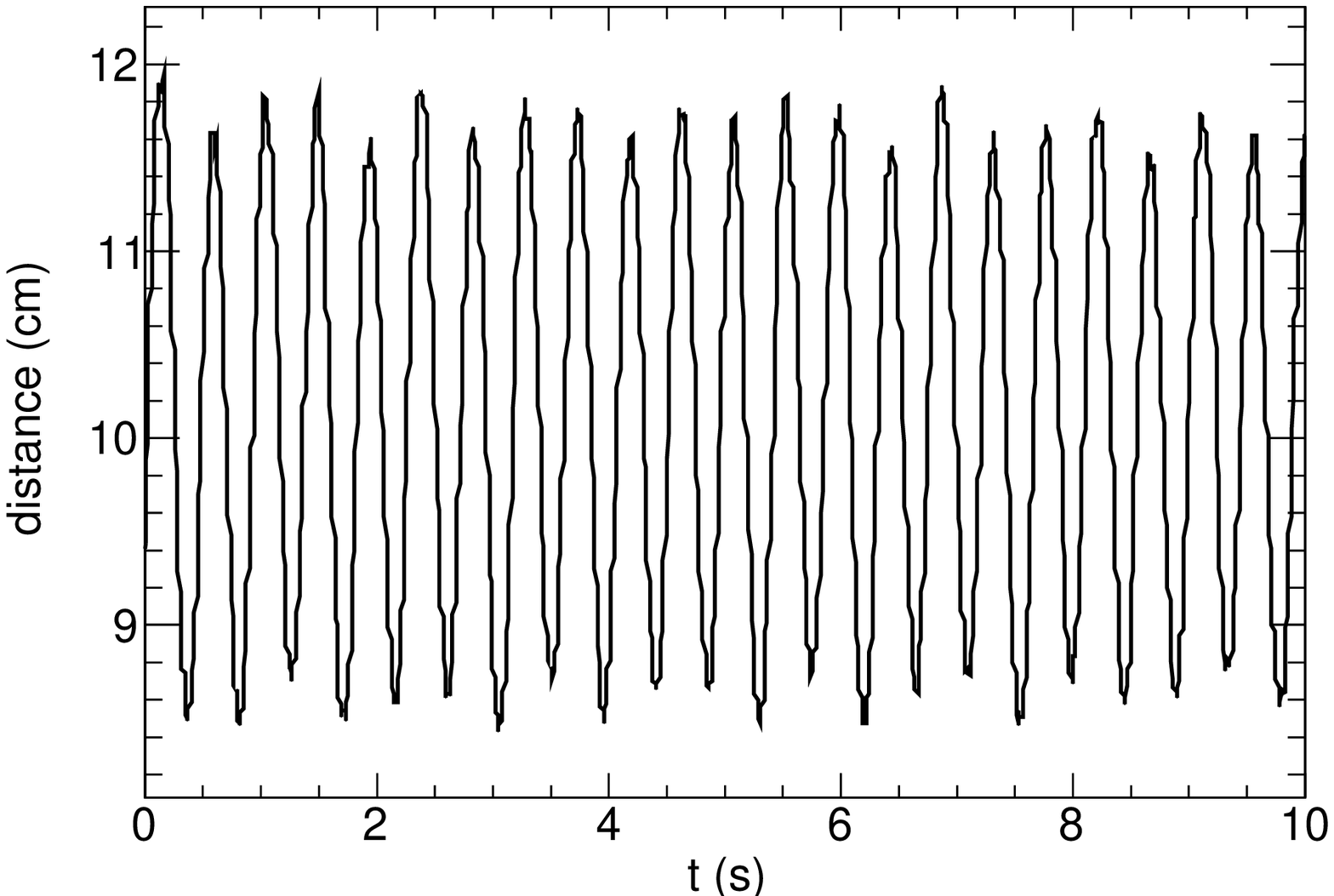}
\includegraphics[width=0.49\textwidth]{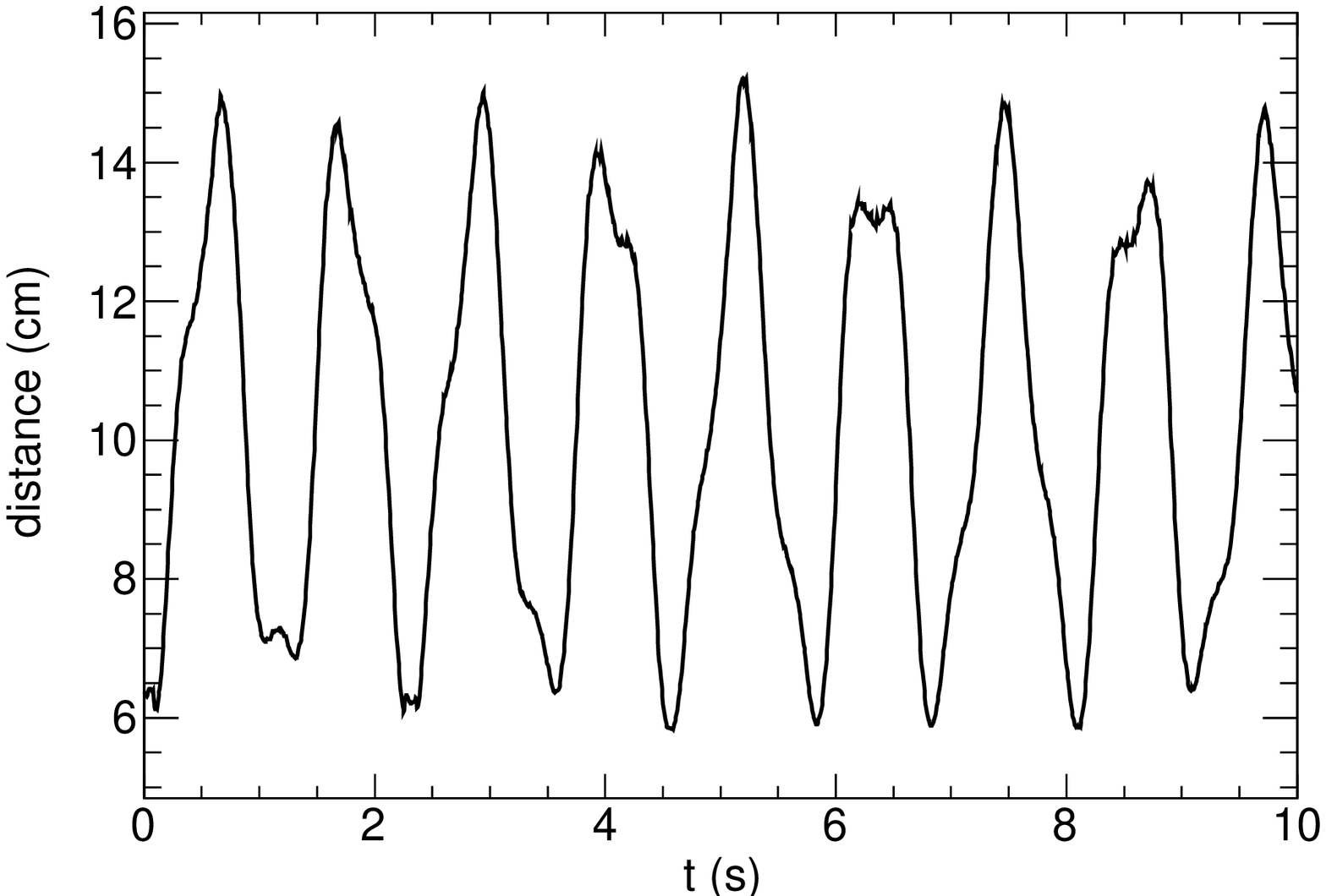}
\caption{Position as a function of time for the lower mass in the coupled spring system. The left panel shows the curve obtained for initial conditions (i) while the right panel represents initial conditions (ii). The same smoothing procedure shown in figure \ref{averagefig} was used here.} \label{coupleddatafig}
\end{figure}

Several configurations for the initial conditions were tested and the resulting displacements of the lower mass for two particularly interesting cases are shown in figure \ref{coupleddatafig}. In all cases, the two springs used in the previous section were coupled vertically to one another and the attached masses were measured to be $\rm m_1 = 100.00(5)~g$ and $\rm m_2 = 95.00(5)~g$. The description of the system and initial configurations are as follow: (i) the two masses were pulled apart from each other (initial displacement of both out of phase) and released; (ii) the mass located between the two springs was put to oscillate while holding the mass at the bottom. A few seconds after the release of the first one, the mass at the bottom was also released and its position as a function of time was recorded.

\begin{figure}[h!]
\centering
\includegraphics[width=0.49\textwidth]{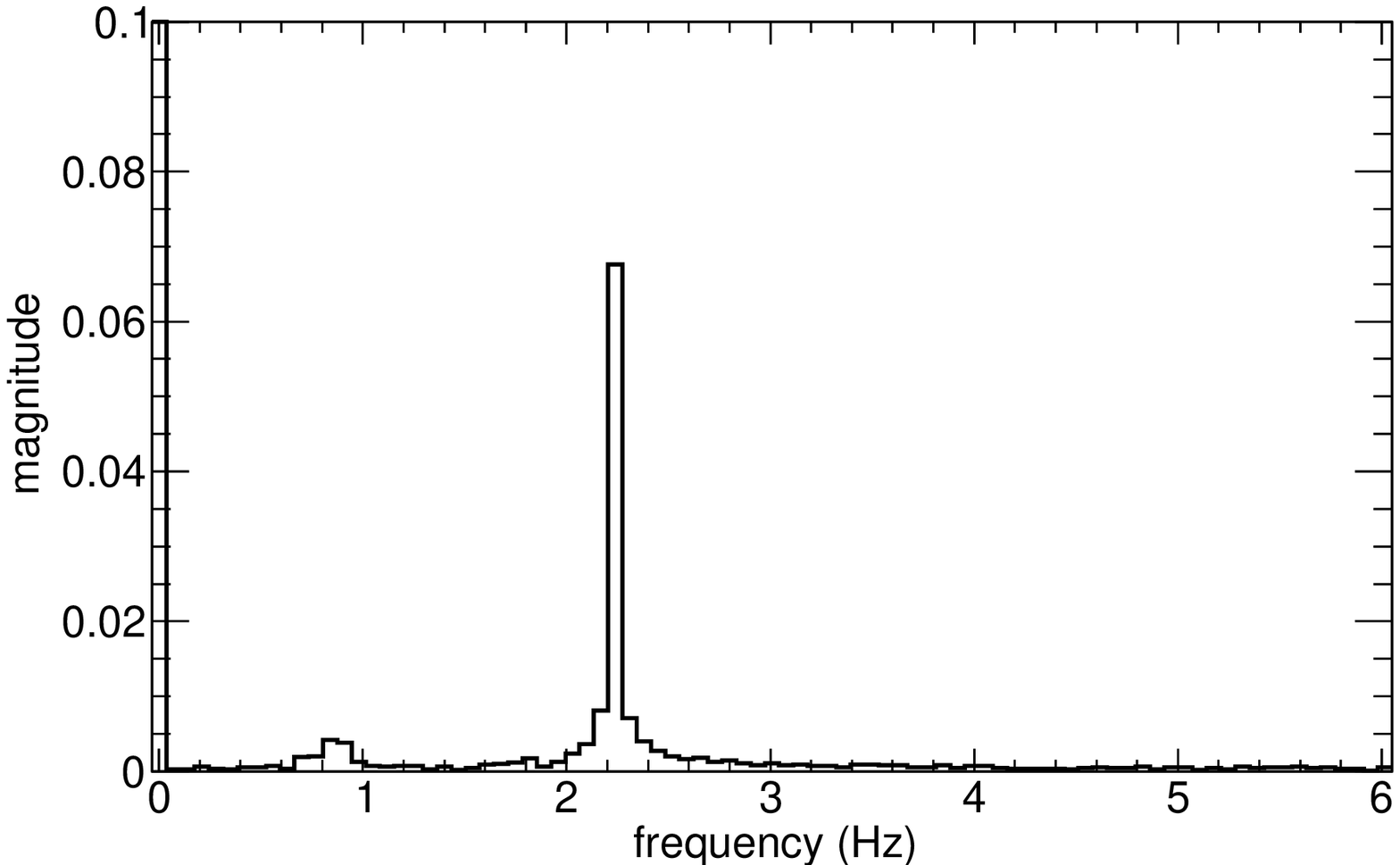}
\includegraphics[width=0.49\textwidth]{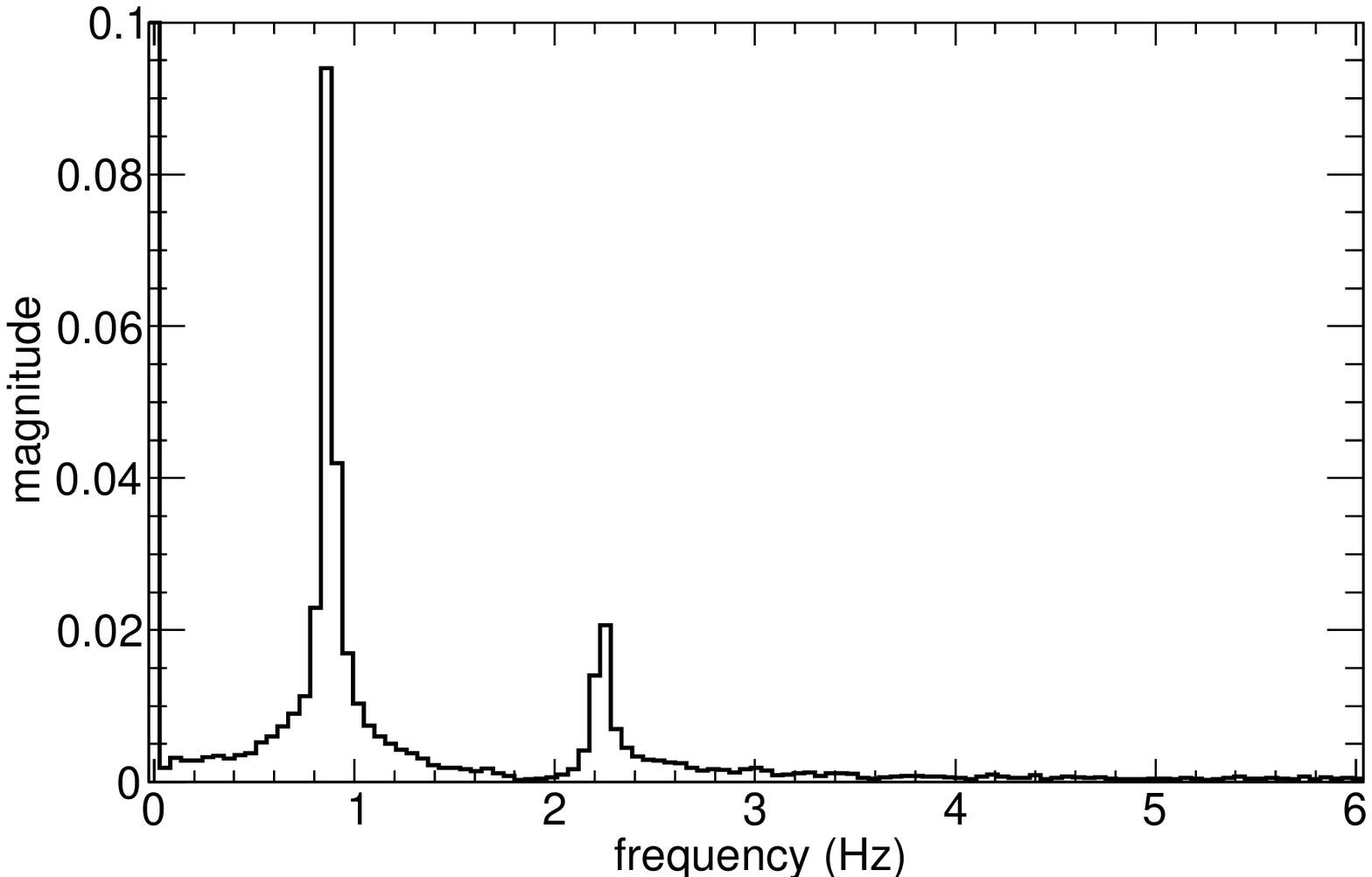}
\caption{Fourier transform for oscillations of a coupled mass-spring system. Initial conditions (i) are shown on the left panel and (ii) on the right panel. The two fundamental oscillation frequencies are clearly seen as well as the different relative intensities that help explain the oscillatory behaviour.} \label{fourier}
\end{figure}

By performing a Fourier transform of the signals seen in figure \ref{coupleddatafig}, one can obtain the two fundamental oscillation frequencies for the coupled springs system (\ref{omegacoup}). The transforms for the data with initial conditions (i) and (ii) are shown in figure \ref{fourier}. The fundamental frequencies found are 0.867(18) Hz and 2.248(34) Hz in accordance with the theoretical values. The Fourier component at 0 Hz gives the equilibrium position for the second mass. For initial condition (i) it can be seen that the highest frequency is very much dominant while for (ii) both frequencies are clearly visible with the lower component amplitude being larger than that of the higher component. It explains the different behaviour of the curves shown in figure \ref{coupleddatafig}.

In a typical classroom the fast Fourier transform can be conveniently performed using SciDavis \cite{scidavis}, for example, which is a user friendly open source software option for data analysis.

\section{Prototype Operation}

The feasibility for the use of the detecting prototype by undergrad students was partially evaluated by assigning a small group of three second year engineering students the task to replicate part of the experimental procedures described in this article. The main goal was to verify whether or not those were interesting activities providing a potential learning experience from the students point of view and if those would be adequate activities without consuming too much time. No precise instructions were provided but only general directions and information on the use of the sensor and microcontroller. 

In particular the students were able to replicate simplified versions of the time and position calibrations. They decided to count the amount of points sampled within a full period of a square wave provided by a signal generator to estimate the sampling interval, a procedure presenting some caveats that can lead to a poor determination of the time interval if not performed correctly. Moreover, it seems that the lack of experience in statistical data analysis induced them to analyse just a single period of the squared wave instead of using several periods for a more accurate and precise determination. This procedure introduced a systematic error leading to a lower sampling time and with higher uncertainty. However, the sampling time can be obtained with precision by simply counting the number of measurements setting the Arduino code to run within a fixed time window.
On the sensor's position calibration the data acquisition was an easy task, however it seemed natural to the students to propose polynomial fits to describe the relationship between measured voltage and distance dividing the whole operation range into three distinct fit regions. At the end satisfactory fits were also obtained.  

The simple and damped harmonic oscillator experiments were also replicated by the students but they reported that for some data acquisitions the curve would come out unexpected or with poor quality if minimal experimental conditions were not met. In such cases they noticed that allowing natural light in, poor sensor-target alignment, or air flow perturbations due to air conditioning in the room
could be causes for data quality degradation. For instance, the setup location had to be such that no direct sun light would hit the sensor, therefore, not affecting the measurements. They also noted that the sensor and the target should be well aligned such that the infrared radiation received by the sensor would be from the reflection on the target only and not residual reflections on other components of the system, providing reliable measurements. Direct strong airflow from the air conditioning also proved to be a possible source for mechanical lateral displacement and shaking of the target. Both effects can cause non periodic disturbances in the readings similar to the other identified sources such that the overall results are bumps or sudden jumps in the position curve as a function of time. Nevertheless, when acquiring data as shown in sections \ref{setup} and \ref{app}, care was taken in order not to have undesirable external influences such that no unexpected effects were seen and it was possible to reproduce the experiments in several different occasions. 

Note that depending on the time constraints or goals for a laboratory activity or teaching class one might want to just focus on studying the physical principles of interest from the kinematic measurements and facilitate its operation giving further instructions instead of leaving the students to figure out how to improve data quality reducing influencing external conditions that could affect the experiment, converting measured voltages for position, etc.  

\section{Conclusion}

We have proposed means to assemble, calibrate and evaluated the performance of a simple to use and inexpensive detection system to make kinematic measurements that can be safely operated in the physics laboratory environment. It has the capability to perform fast data acquisition, $\approx 400 \mathrm{\mu s}$ between measurements, enough for its application on the study of mass-spring systems, for example. The prototype allows to exploit the typical features oscillating systems present through the detailed observation of the time evolution of such systems providing good quality estimates for periods, natural frequencies and damping. More complicated setups can be easily built and analysed such as a coupled pair of mass-spring systems for which its initial conditions dictate the overall movement behaviour.

We must also point out that there is a distance limitation for the particular sensor used here, providing good estimates in a $\sim 25$ cm range. For recording wider movements, other sensors from the SharpGP family could be suitable for the $\rm 10-80~cm$ and $\rm 20-150~cm$ distance ranges\cite{sharp2}, for example. In this case, only new position calibrations should be performed but the general idea on how to assemble and use the experimental set is essentially the same.

A typical dispersion on the measured voltage was observed to be lower than 9\% across the investigated range. Nevertheless, to limit the influence of statistical fluctuations and to improve position estimates an averaging procedure was adopted at the data analysis level avoiding more elaborate model based statistical fits or the use of a hardware integrator before data transfer keeping the electronics as simple as possible. However, these are extra techniques that could be incorporated and may be suitable for undergrad classes. 

As already mentioned not only the authors but undergrad students were able to use this setup for oscillatory experiments. We point out that several tests were carried out at different times by different operators. In all occasions, once the procedures described in this article were followed, reproducibility of the position curves was achieved. The fact that we could also obtain a dynamic calibration curve with a non periodic motion indicates that this apparatus is well suited for other applications.

The interesting feedback provided by students makes it clear that the prototype is a viable device for physics laboratories. Only very simple measures are necessary to remove external influences on the experiments in order to isolate the desired phenomenon to be analysed. If such precautions are taken, this device has been shown to provide excellent results with its simplicity and easy to use. It is relatively low cost and the user has total control of the data taking and analysis.

\ack{The authors acknowledge support from CNPq (grant 481953/2012-4) and would like to thank the students A. Lippi, B. Medeiros and P. Serrano for participating in some of the learning activities proposed in this work.}

\end{document}